\documentclass[twocolumn]{revtex4}
\usepackage{amsmath,amsfonts,epsfig,graphicx,float}

\begin{document}

\title{Complex contact-based dynamics of microsphere monolayers revealed by resonant attenuation of surface acoustic waves}

\author{M. Hiraiwa$^{1}$, M. Abi Ghanem$^{1}$, S. P. Wallen$^{1}$, A. Khanolkar$^{1}$, A. A. Maznev$^{2}$, and N. Boechler$^{1}$}

\affiliation{ 
$^1$ Department of Mechanical Engineering, University of Washington, Seattle, WA 98195 \\
$^2$ Department of Chemistry, Massachusetts Institute of Technology, Cambridge, MA 02139 
}



\begin{abstract}

Contact-based vibrations play a critical role in the dynamics of granular materials. Significant insights into vibrational granular dynamics have been obtained with reduced-dimensional systems containing macroscale particles. We study contact-based vibrations of a two-dimensional monolayer of micron-sized spheres on a solid substrate. Measurements of the resonant attenuation of laser-generated surface acoustic waves reveal three collective vibrational modes involving both displacements and rotations of the microspheres. To identify the modes, we tune the interparticle stiffness, which shifts the frequency of the horizontal-rotational resonances while leaving the vertical resonance unaffected. From the measured contact resonance frequencies we determine both particle-substrate and interparticle contact stiffnesses and find that the former is an order of magnitude larger than the latter. This study paves the way for investigating complex contact-based dynamics of microgranular media, demonstrates a novel acoustic metamaterial, and yields a new approach to studying micro- to nanoscale contact mechanics in multiparticle networks.

\end{abstract}

\maketitle


Micro- and nanoscale particles in contact with other bodies experience strong adhesive forces that induce deformation near the point of contact \cite{IsraelachviliBook}. The understanding of contact mechanics is critical to many fields, including areas such as surface science \cite{IsraelachviliBook}, contaminant removal \cite{MittalBook}, self-assembly \cite{Israelachvili2008}, powder technology and processing \cite{DuranBook,PowderTechBook}, and biomedicine \cite{BioMed}. In systems with adhered micro- and nanoscale particles, low frequency dynamic disturbances (compared to the intrinsic spheroidal modes of the spheres \cite{Sato1962}) can induce contact-based vibrational modes in single- and multi-particle systems, where the particles move like rigid bodies and the local region of deformation around the contact acts as a spring \cite{NesterenkoBook}. 

Such contact-based vibrational modes form the foundation for the dynamics of particulate assemblies. The contact-based dynamics of granular media play a critical role in fields such as wave propagation in geological and other microstructured materials \cite{NesterenkoBook}. While there has been significant progress in the study of the contact-based dynamics of macroscale granular media \cite{NesterenkoBook,GranularCrystalReviewChapter,Jamming}, the dynamics of micro- to nanoscale particle assemblies are less understood. This difference in scale is important from a fundamental perspective; in particular, adhesion forces negligible for macroscale particles become critical at micro- to nanoscales. 

At the macroscale, studies of reduced dimensional systems, such as one- and two-dimensional granular arrays, have yielded significant insights into the dynamics of granular materials \cite{NesterenkoBook,GranularCrystalReviewChapter}. In contrast, studies of the contact-based dynamics of micro- to nanoscale particle assemblies have hitherto been restricted to three-dimensional, typically disordered, settings \cite{NesterenkoBook,Fytas2012, Fytas2014, Ayouch2012}. Recently, a contact resonance of microspheres assembled into a two-dimensional monolayer adhered to a solid substrate was measured via its hybridization with surface acoustic waves (SAWs) traveling in the substrate \cite{Boechler2013}. The results agreed well with a simple model where the particle motion was restricted to the vertical (out-of-plane) degree of freedom and the interaction between the particles was disregarded. However, more advanced models \cite{Wallen2015,Gusev2011} predicted complex dynamics that are strongly affected by particle rotations. For motion in the sagittal plane, a close-packed monolayer of spheres on a solid substrate is expected to yield three collective contact-based vibrational modes: one predominantly vertical, and two of mixed horizontal-rotational character, all of which should interact with SAWs \cite{Wallen2015}. It has remained a mystery as to why the previous experiment only showed the presence of a single contact resonance mode.

\begin{figure*}[t]
\begin{center}
\includegraphics[width=18 cm]{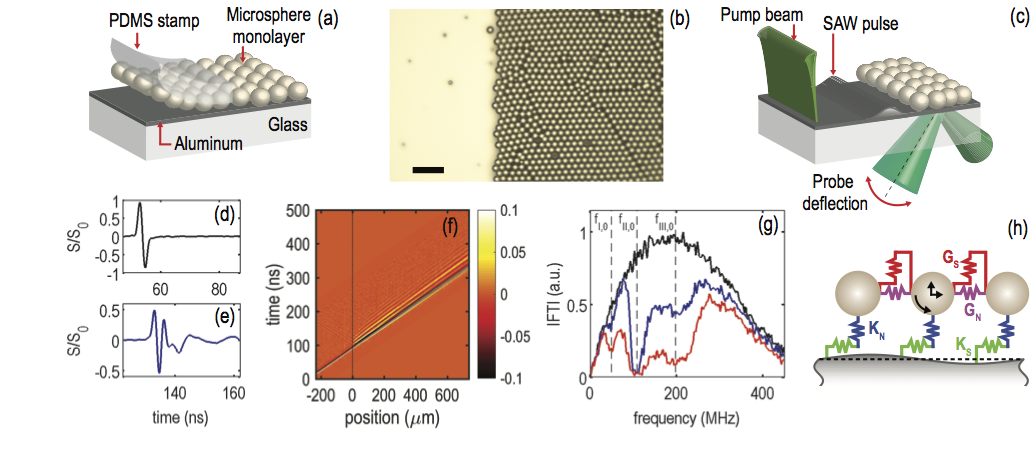}
\end{center}
\caption{\label{Figure1} [Color online] Overview of the experiment. (a) PDMS stamp is used to remove a portion of a microsphere monolayer deposited on aluminum-coated glass microscope slide. (b) Microscope image of the interface between monolayer and blank sample regions. The scale bar is 10 $\mu$m. (c) Schematic of the scanned laser ultrasonic experimental setup. (d) Normalized signal measured in the blank region. (e) Normalized signal measured $132$ $\mu$m inside the monolayer region. (f) Spatiotemporal plot of the normalized measured signals. Position denotes distance from the interface. The vertical dotted line denotes the interface. (g) Normalized Fourier spectra of the signals in (d) and (e) using the same colors. The red curve is the spectrum of a signal measured $400$ $\mu$m inside the monolayer region. Vertical dashed lines denote the identified contact resonance frequencies. (h) Schematic of the dynamical model.}
\end{figure*}
  
In this work, we reveal the presence of all three contact resonances predicted for the microsphere monolayer, by measuring the attenuation of SAWs using a scanned laser ultrasonic technique. We test the model by changing the interparticle contact stiffness via deposition of a thin aluminum film on top of the spheres, which shifts the horizontal-rotational contact resonance frequencies upwards. We further confirm the nature of the modes using a complementary laser-ultrasonic technique that preferentially excites the vertical contact resonance. In addition to providing direct evidence of the rotational-vibrational dynamics of microgranular media, our work opens a new approach for the study of micro- to nanoscale particle contact mechanics by enabling measurements of both interparticle and particle-substrate contact stiffness and offering insight into the role of shear contact rigidity.

\section{Results}
%
Our sample is composed of a monolayer of $D$ = 2.0 $\mu$m diameter silica microspheres deposited on an aluminum-coated glass substrate, as shown in Figure~\ref{Figure1}a,b. The aluminum layer is $100$ nm thick, and the glass is $1.5$ mm thick. A wedge-shaped cell convective self-assembly technique is used to assemble the monolayer on the substrate \cite{WedgeCell}.  To obtain a planar interface between substrate regions with and without the microsphere monolayer (hereafter referred to as the monolayer and blank regions, respectively), we use a micro-contact-printing method, as shown in Fig.~\ref{Figure1}a, wherein a soft Polydimethylsiloxane (PDMS) stamp is pressed into conformal contact with the microsphere monolayer and then is removed, such that the spheres detach from the substrate in the stamped region \cite{muCP}. A representative optical microscope image of the resulting interface is shown in Fig.~\ref{Figure1}b.

To generate and measure SAW propagation in our sample, we utilize a scanned laser-ultrasonic pump-probe technique, as shown in Fig. \ref{Figure1}c. We focus a sub-nanosecond laser pulse, which serves as a ``pump,'' into a line on the aluminum surface of the blank region of the substrate. The absorbed laser light induces rapid thermoelastic expansion of the aluminum film that launches counter-propagating, broadband SAW pulses in the substrate. Because of the high aspect ratio of the pump spot, the SAW pulse propagates as a plane wave in the direction perpendicular to the line source. The acoustic response of the sample is measured via a knife-edge photo-deflection technique \cite{ScrubyBook}.  A continuous-wave ``probe'' beam, is incident through the substrate and focused to a small spot on the aluminum film. The reflected probe light is focused onto a fast photodetector, after being partially blocked by a sharp knife-edge. Changes in surface slope and refractive index in the sample caused by the propagating SAWs deflect the probe beam, which translates to a change in intensity on the photodetector. To obtain spatial information, the sample is scanned in the direction of the SAW propagation using a motorized translation stage. Both pump and probe are initially focused onto the blank region. The pump and probe remain at a fixed relative distance throughout the experiment. As the sample is scanned, both pump and probe move progressively closer to the interface, with the probe crossing into the monolayer region. The scan is stopped just before the pump reaches the interface. 

%

\begin{figure*}[t]
\begin{center}
\includegraphics[width=18cm]{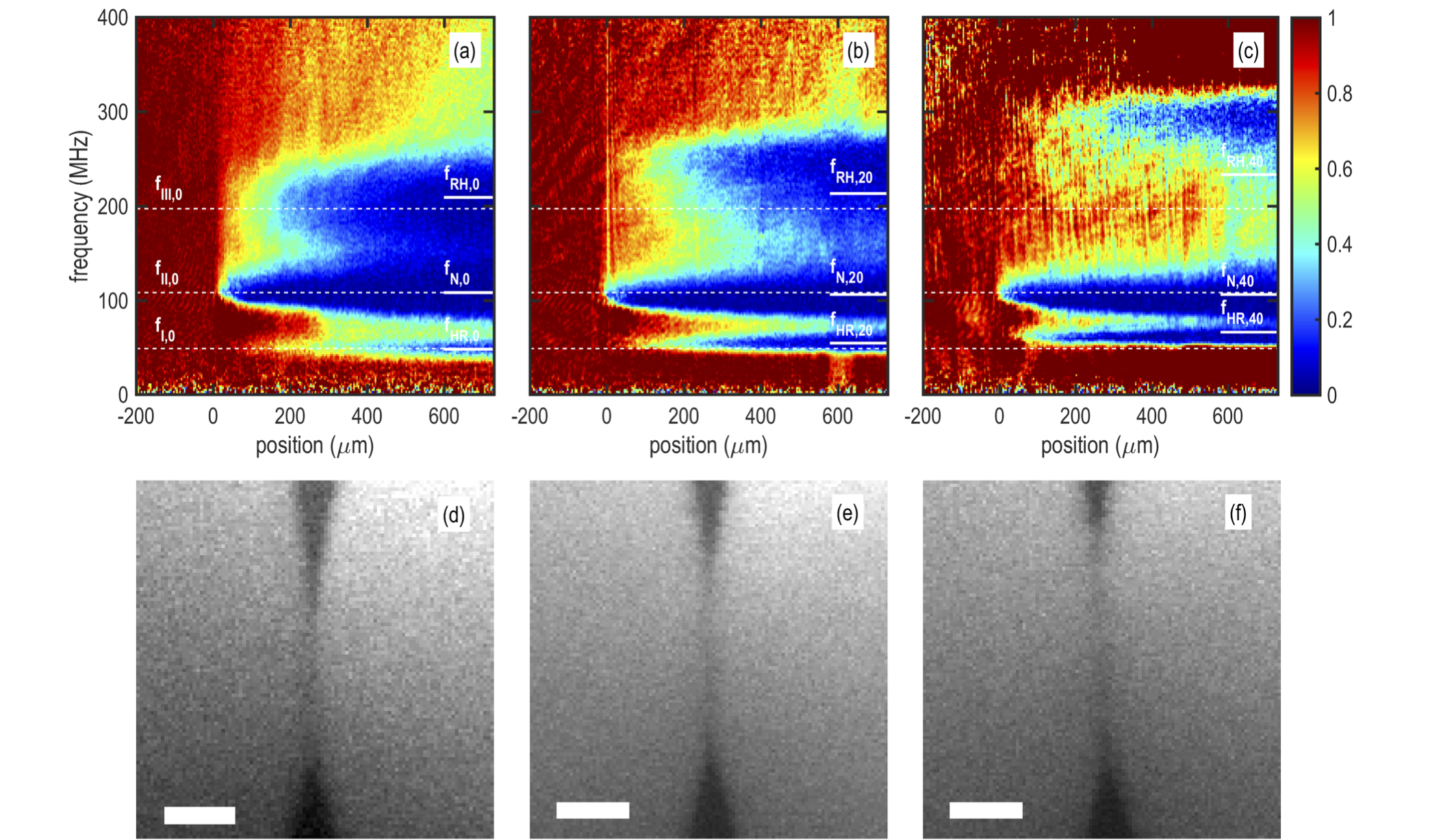}
\end{center}
\caption{\label{Figure3} [Color online] (a-c) Transmission spectra for SAWs propagating across the interface between blank and monolayer regions. The color bar denotes the magnitude of the transmission coefficient. Horizontal dashed lines denote the identified contact resonance frequencies for the uncoated monolayer. Short horizontal lines on the right of the panels are the fitted contact resonance frequencies. The second number in the subscript of the contact resonance frequencies denotes the thickness of the aluminum coating in nanometers. Position denotes distance from the interface. (a) Uncoated microsphere monolayer. (b) $20$ nm of aluminum coating. (c) $40$ nm of aluminum coating. (d-f) SEM images of the same interparticle contact, corresponding to the spectra of (a-c), respectively. The scale bar is $100$ nm.}
\end{figure*} 

Figure~\ref{Figure1}d,e shows typical measured signals $S$, normalized to the maximum  signal amplitude $S_0$ measured during the scan. Figure~\ref{Figure1}d corresponds to a probe position in the blank region, and Fig.~\ref{Figure1}e corresponds to a probe position 132 $\mu$m inside the monolayer region. The distortion of the signal in Fig.~\ref{Figure1}e, with respect to Fig.~\ref{Figure1}d, is a result of dispersion and dissipation induced by the monolayer. Figure~\ref{Figure1}f presents a spatiotemporal plot of the signals measured throughout the scan, which shows the distortion of the pulse as it propagates through the monolayer region. 

The Fourier spectra of the normalized signals in Fig. ~\ref{Figure1}d and Fig.~\ref{Figure1}e are shown in Fig.~\ref{Figure1}g. The spectrum corresponding to the signal in the monolayer region shows a sharp dip at $108$ MHz. We also observe two smaller dips surrounding this resonance, and denote the three dips with vertical lines drawn at $f_{I,0}=49$ MHz, $f_{II,0}=108$ MHz, $f_{III,0}=197$ MHz. We also show a third spectrum, corresponding to a location $400$ $\mu$m inside the monolayer region, which demonstrates the evolution of the attenuation zones. 

To obtain position-dependent transmission spectra of SAWs traversing the interface, we normalize the Fourier spectra at each position by the average Fourier spectra of the incident SAW (averaged over all positions in the blank region). Figure~\ref{Figure3}a shows the measured transmission spectra as a function of distance from the interface. In the transmission spectra of Fig.~\ref{Figure3}a, three distinct attenuation maxima are evident, corresponding to the identified dips in Fig.~\ref{Figure1}g. We interpret the measured attenuation maxima as being caused by the interaction of SAWs with contact resonances of the microsphere monolayer, as described by the recently developed model of Ref. \cite{Wallen2015}. In this model, the microspheres are considered as rigid bodies, and the sphere-substrate and sphere-sphere contacts are represented as normal and shear springs, as is shown in Fig.~\ref{Figure1}h. This model predicts three collective vibrational modes of the monolayer involving vertical, horizontal, and rotational motion of
spheres in the sagittal plane.

At long wavelengths (compared to the particle spacing), one of the modes is purely vertical, with a frequency given by 
\begin{align}
\begin{split}
f_N & = \frac{1}{2\pi} \left[\frac{K_N}{m}\right]^{1/2},\\
 \label{Res_longwave1}
\end{split}
\end{align}
\noindent
while two others are of mixed horizontal-rotational character, with frequencies given by
\begin{align}
\begin{split}
f_{RH} & = \frac{1}{2\pi} \left[\left(\frac{K_S}{4m}\right) \left(20 \gamma + 7 + \sqrt{400 \gamma^2 + 120 \gamma + 49}\right)\right]^{1/2}\\
f_{HR} & = \frac{1}{2\pi} \left[\left(\frac{K_S}{4m}\right) \left(20 \gamma + 7 - \sqrt{400 \gamma^2 + 120 \gamma + 49}\right)\right]^{1/2}, \label{Res_longwave2}
\end{split}
\end{align}
\noindent
where $m=\rho \pi D^3 /6$ is the microsphere mass, $K_N$ is the particle-substrate normal stiffness, $K_S$ is the particle-substrate shear stiffness, $G_S$ is the interparticle shear stiffness, and $\gamma=G_S/K_S$. The interparticle normal contact stiffness $G_N$ does not affect these resonances at long wavelengths. The frequency $f_{RH}$ corresponds to the predominantly rotational mode and is always higher than the frequency of the predominantly horizontal mode $f_{HR}$. If the monolayer is placed on an elastic substrate, all three modes are predicted to interact with SAWs \cite{Wallen2015}. In the absence of dissipation, this interaction results in the hybridization and avoided crossing of the Rayleigh SAW with the contact resonances. In the presence of dissipation, avoided crossing may or may not take place, but one would invariably expect a peak in attenuation at the contact resonance frequency \cite{Garova1999}.  As can be seen from Eq.~\ref{Res_longwave1} and Eq.~\ref{Res_longwave2}, $f_{N}$ is determined solely by the particle-substrate contact, whereas $f_{RH}$ and $f_{HR}$ are affected by both contacts. Hence, if we increase the interparticle contact stiffness, only $f_{RH}$ and $f_{HR}$ are expected to increase.

To test the model and verify the nature of the observed contact resonances, we coat the microsphere monolayer with a thin aluminum layer, which stiffens the interparticle contact without affecting the particle-substrate contact. Figures~\ref{Figure3}d-f show scanning electron microscope (SEM) images for the same interparticle contact in the uncoated sample, and after coating it with $20$ nm and $40$ nm of aluminum. Additional SEM images also confirm that the particle-substrate contact is shaded from the aluminum deposition by the microsphere. Figure~\ref{Figure3}b and Fig.~\ref{Figure3}c show transmission spectra for the samples coated with aluminum. We clearly see that the highest and the lowest attenuation maxima shift upwards upon the deposition of the aluminum, while the middle maximum remains nearly unaffected. The relatively small downshift of the middle resonance, which is approximately consistent with the predicted frequency downshift of $\sim4$\% due to extra mass loading, confirms our assignment of the middle resonance to $f_N$. In all cases, the middle zone has the largest attenuation, indicating stronger coupling of this resonance to the propagating SAWs. This stronger coupling suggests that this is the same resonance observed in previous studies \cite{Boechler2013,Khanolkar2015}. 

\begin{figure}[t]
\begin{center}
\includegraphics[width=8.5 cm]{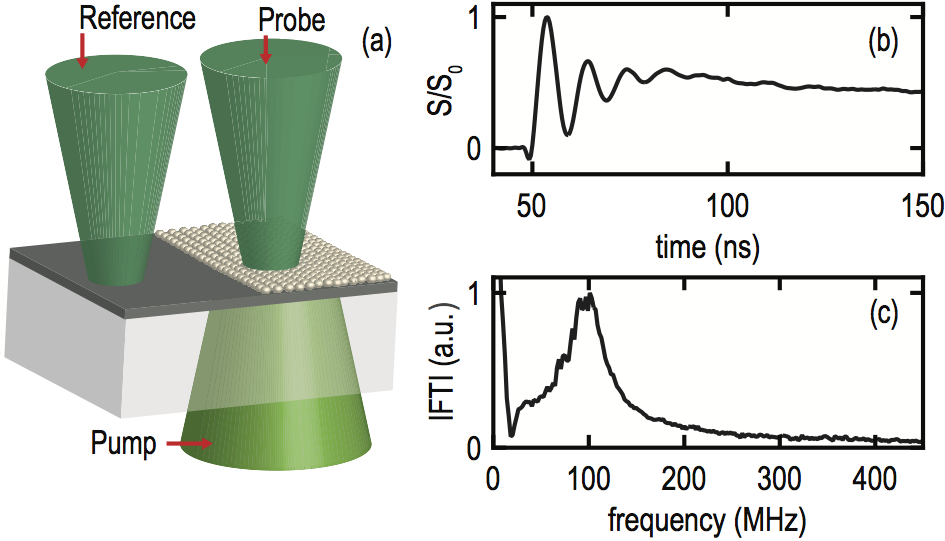}
\end{center}
\caption{\label{Figure5} [Color online] (a) Schematic of the experiment with large spot excitation and grating interferometer detection. The interferometer uses a reference beam reflected from a blank (and unperturbed) region of the sample. (b) Normalized signal measured with the grating interferometer. (c) Fourier spectrum of the signal in (b). }
\end{figure}

For further confirmation of the assignment of the resonances, we conduct a separate experiment on the sample coated with $40$ nm of aluminum,wherein a pump beam entering through the substrate is focused to a large diameter ($240$ $\mu$m at $1/e^2$ intensity level) spot, as shown in Fig.~\ref{Figure5}a. In this configuration, thermal expansion of the aluminum layer excites the vertical contact resonance of the spheres directly, while horizontal-rotational resonances are not excited because of symmetry constraints. The displacement of the spheres is measured with a grating interferometer \cite{NelsonGrating}, which is also only sensitive to vertical motion. The measured signal shown in Fig. \ref{Figure5}b contains oscillations at a frequency of $\sim100$ MHz, as can be seen from the Fourier spectrum in Fig.~\ref{Figure5}c. Thus the middle resonance in Fig.~\ref{Figure3}a-c is identified as the vertical mode. 

We compare the frequencies of the observed attenuation maxima shown in Fig.~\ref{Figure3} with those predicted by Eq.~\ref{Res_longwave1} and Eq.~\ref{Res_longwave2}. While the equations have three unknown parameters ($K_N$, $K_S$, and $G_S$), we relate $K_S$ to $K_N$ via the Hertz-Mindlin contact model \cite{Mindlin}, which leaves two independent parameters. In the Hertz-Mindlin contact model, assuming a no-slip condition at the contact, the normal stiffness for a given contact is related to the shear stiffness, such that $K_S/K_N=4G^{*}/E^{*}$, where $E^* = [(1-\nu_1^2)/E_1 + (1-\nu_2^2)/E_2]^{-1}$ is the effective Young's modulus of the contact, and $G^* = [(2-\nu_1)/G_1 + (2-\nu_2)/G_2]^{-1}$ is the effective shear modulus, where $E_1$ and $G_1$ are the moduli for the silica microspheres, and $E_2$ and $G_2$ are the moduli for the aluminum substrate \cite{Supplementary}. Using Eq.~\ref{Res_longwave1} and the measured value of $f_{II,0}=f_N$, we find a particle-substrate normal contact stiffness of $K_N=4.0$ kN/m, and thus also obtain the particle-substrate shear stiffness $K_S=3.5$ kN/m. We then use a least-squares fit to determine the interparticle shear stiffness $G_S$, where the quantity $((f_{I}-f_{HR})/f_{I})^2+((f_{III}-f_{RH})/f_{III})^2$ is minimized, with $f_{HR}$ and $f_{RH}$ defined as in Eq.~\ref{Res_longwave2}. For the uncoated sample, we obtain an interparticle shear stiffness of $G_S=0.3$ kN/m \cite{Supplementary}. In Fig.~\ref{Figure3}, we denote the fitted contact resonance frequencies using white solid lines on the right side of each panel. For the uncoated sample, we see an excellent agreement between the measured attenuation frequencies and the fitted contact resonance frequencies. For the sample coated with $40$ nm of aluminum, the agreement is not as good. This difference may be due to deviations from the physical scenario described by our model due to the presence of the aluminum, including asymmetric interparticle contacts and the formation of ``necks'', which may lead to bending resistance not taken into account in the model.

\begin{figure}[t]
\begin{center}
\includegraphics[width=9 cm]{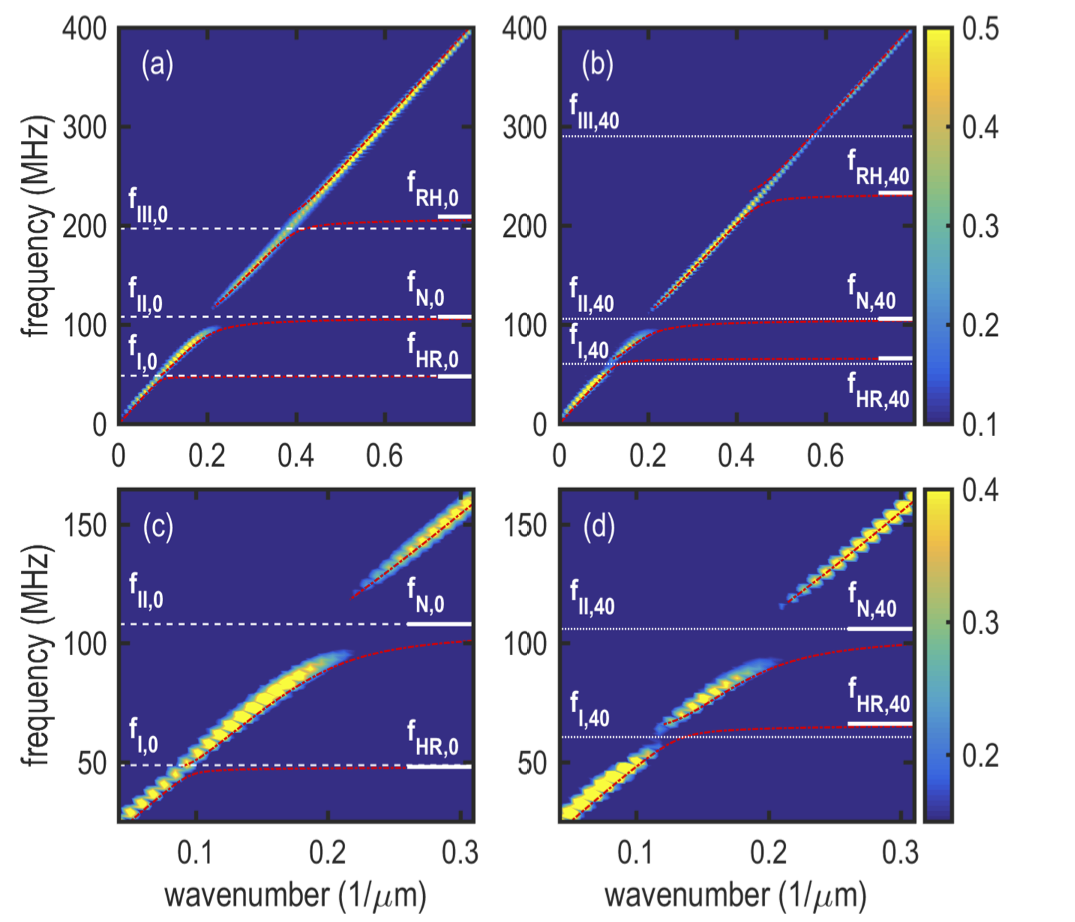}
\end{center}
\caption{\label{Figure4} Surface acoustic wave dispersion in samples with an (a,c) uncoated monolayer and (b,d) $40$ nm thick aluminum coating. The color plot denotes the calculated 2D Fourier spectra of SAWs propagating in the sample, and the color bar the normalized magnitude. Horizontal dashed and dotted lines correspond to the identified contact resonance frequencies for the uncoated sample and the sample coated with $40$ nm of aluminum, respectively. Short horizontal lines on the right of the panels are the fitted contact resonance frequencies. The second number in the subscript of the contact resonance frequencies denotes the thickness of the aluminum coating in nanometers. Red dash-dotted lines are the dispersion curves calculated using the fitted resonance frequencies. Panels (c) and (d) provide a closer view of the lower part of (a) and (b), respectively.}
\end{figure}

We have also studied the effect of the microspheres on SAW dispersion. Figure~\ref{Figure4} shows the normalized magnitude of the two-dimensional (space-time) Fourier transform of the scanned measurements taken in the monolayer region. Figure~\ref{Figure4}a,c shows spectra corresponding to the uncoated sample. Figure~\ref{Figure4}b,d shows spectra corresponding to the sample with $40$ nm of aluminum. The spectra show a line corresponding to Rayleigh SAWs in the substrate, which has three gaps or regions of attenuation corresponding to the attenuation zones seen in Fig~\ref{Figure3}. The highest and lowest zones appear as lighter-colored, attenuated regions, whereas the middle gap shows a clear gap with curvature indicative of an avoided crossing. An emerging band gap can also be seen at the lowest resonance in Fig.~\ref{Figure4}b,d. 

Using the fitted contact resonance frequencies with the effective medium model for a monolayer of microspheres on an elastic substrate~\cite{Wallen2015}, we plot the calculated dispersion curves as the red dash-dotted lines in Fig.~\ref{Figure4} \cite{Supplementary}. For the strong middle resonance and also the emerging avoided crossing at the lowest resonance in Fig.~\ref{Figure4}b,d, we see reasonable agreement between experiment and theory in the curvature of the branches, which confirms that our model captures the coupling strength between the contact resonances and the SAWs. For resonances with weak coupling and low quality factors, hybridization gaps can appear as attenuation zones instead of avoided crossings \cite{Garova1999}. Thus weak, broad resonances seen in the attenuation data are not necessarily discernible in the dispersion measurements; this explains why weaker horizontal-rotational resonances were not identified previously in Ref. \cite{Boechler2013}. Another factor that makes the current experiment more sensitive is a longer SAW propagation distance involved, and greater resolution in the wave vector domain.  

\section{Discussion}

A monolayer of microspheres on a substrate can be considered as a locally resonant metamaterial \cite{Liu2000,Achaoui2013} for surface acoustic waves. Acoustic metamaterials are typically studied in the context of designing structures with tailored effective properties for controlling the propagation of acoustic waves. In this study, we measured acoustic waves propagating through the metamaterial to explore the dynamics of microgranular media, just as optical spectroscopy is used to study dynamics of atomic or molecular assemblies. Our study reveals the critical role of particle rotations: for instance, without rotations, the upper ($RH$) resonance would not be present \cite{Wallen2015}. Furthermore, while previously studied locally resonant acoustic metamaterials typically involved isolated resonant inclusions, the horizontal-rotational resonances of a granular monolayer revealed in this study involve collective dynamics of the microspheres. The observed contact resonances offer insight into the contact mechanics of microspheres by enabling the simultaneous measurement of interparticle and particle-substrate contact stiffnesses. 

The most intriguing result is that the interparticle shear contact stiffness is over an order of magnitude smaller than the particle-substrate  contact stiffnesses. As a comparison, we use the Hertz-Mindlin elastic contact model in combination with the DMT model of an adhesive contact \cite{DMT} to calculate theoretical contact stiffnesses. This results in predicted stiffnesses $K_{N,DMT}=1.6$ kN/m, $K_{S,DMT}=1.4$ kN/m, and $G_{S,DMT}=0.8$ kN/m, which gives a ratio of less than two between the particle-substrate and interparticle stiffnesses. The measured particle-substrate normal contact stiffness is over twice as large as predicted, consistent with the trend observed in recent studies \cite{Boechler2013,Khanolkar2015}. A discrepancy between estimated and measured values can be ascribed to factors such as uncertainty in the work of adhesion \cite{IsraelachviliBook,Supplementary}, plastic deformation, which may stiffen the contact \cite{SilicaNanospherePlasticity,AgingPRL}, or microslip at the contact, which may decrease the shear contact stiffness \cite{Johnson}. However, these factors are unlikely to fully explain the magnitude of the observed disparity between interparticle and particle-substrate contact stiffnesses. An examination of SEM images \cite{Supplementary} showed that the interparticle contact network is not uniform: even in closely packed regions, most particles do not form adhesive contacts with all six neighbors. This raises the question of how adhesive contact networks form following self-assembly. The ability of our measurements to characterize the average contact stiffness in a multi-particle network offers a unique opportunity to investigate this issue.  

The validation of microscale contact mechanics models \cite{MittalBook} is known to be particularly challenging, with large differences routinely seen between different experiments, and between experiments and theory \cite{Johannsmann2010}. The most common characterization approaches involve the use of atomic-force-microscopes (AFM)  \cite{MittalBook,Fuchs2014} or particle detachment methods \cite{PowderTechBook}. Our method, while being non-contact and non-destructive, offers information about equilibrium contact stiffnesses, and, in contrast to other dynamic techniques involving isolated particles \cite{Cetinkaya2005,Johannsmann2010,Audoin2012}, enables the measurement of interparticle contact stiffness in a microscale multiparticle assembly. 

This work opens the door for the study of the contact-based dynamics of low-dimensional microgranular systems and their underlying contact mechanics. The discovery of collective vibrational modes including rotations as well as displacements, along with the characterization of shear and normal contact stiffnesses, will lead to better understanding of wave propagation in three-dimensional microscale granular media with applications in shock mitigation, energetic materials, seismic exploration, and powder processing. We note that the first evidence of rotational granular dynamics in a macroscale 3D granular medium has just recently been reported \cite{TournatRotationalModes}. Contact-based dynamics in systems of statically-compressed macroscale particles have also been used in the design of nonlinear phononic crystals and metamaterials for stress wave manipulation \cite{GranularCrystalReviewChapter} and signal processing applications \cite{Boechler2011}. Our study points the way towards future micro- to nanoscale analogs, which may have acousto-optic \cite{Bayer2008} or acousto-plasmonic \cite{Plasmonics} functionalities, and be rapidly and inexpensively manufactured via self-assembly. Finally, the sensitivity of the SAW attenuation to contact forces in the microgranular monolayer may be used to make SAW sensors for bioanalytical \cite{Gronewold} and other applications.

\section{Methods}

\noindent {\it Sample fabrication}. 
The aluminum-coated, soda lime glass microscope slide was purchased from EMF Corp. The slide was rinsed with deionized water, acetone and isopropanol, and dried under air flow. It was then hydrophilized in a hydrogen peroxide bath ($80$~$^{\circ}$C) for fifteen minutes. The microspheres were purchased from Corpuscular Inc. as a suspension of $5.0$ wt\% in water. Prior to the deposition, the suspension was further diluted to $1.25$ wt\%, and was ultrasonicated for $3$ minutes. The angle of the wedge formed between the substrate and the top glass slide was $3^{\circ}$.  We deposited $40$ $\mu$L of the diluted suspension into the wedge. The whole setup was then placed on a tilt, such that the meniscus drying front receded upwards on an incline of $15^{\circ}$. The assembly was allowed to dry in ambient lab environment ($40$\% relative humidity, $22^{\circ}$ C). The PDMS stamp was made of Sylgard $184$ with a curing agent to base ratio of 1:10 and left to cure in an oven at ~$70^{\circ}$ C for a day to ensure solidification. The aluminum coating was deposited via electron-beam evaporation. The thickness of the evaporated aluminum layer was measured with a quartz crystal microbalance (QCM). 
\\

\noindent {\it Photoacoustic characterization system}. The pulsed laser source used as the pump has $532$ nm wavelength, $430$ ps pulse duration, energy of 4 $\mu$J per pulse, and $1$ kHz repetition rate. The light was focused to a spot of $1.2$ mm $\times$ $20$ $\mu$m (axis length, at $1/e^2$ intensity level). The resulting surface displacement of the propagating SAW pulse is estimated to be $150$ pm \cite{Boechler2013}. The continuous probe laser has a wavelength $514$ nm and average power of $4$ mW at the sample. The probe passes through the glass substrate and is focused at the aluminum film to a 6 $\mu$m diameter spot (at $1/e^2$ intensity level). A balanced detection scheme is used, in which a metal coated right-angle prism acts as the €˜knife edge for two photodetectors. The separation between the focused pump and the probe beams was ~$850$ $\mu$m. In the grating interferometer experiments, the probe has a diameter of $80$ $\mu$m (at $1/e^2$ intensity level) and a power of $20$ mW at the sample. 
\\

\noindent {\it Spatiotemporal signal processing}. For the spatiotemporal data shown in Fig. ~\ref{Figure1}d-f and Fig.~\ref{Figure4}, we compensate for the fixed pump-probe separation distance by shifting each measured signal in time by $x_s/c_R$ (where $x_s$ is the distance from a reference position, and $c_R=3128$ m/s is the Rayleigh wave speed in the aluminum-coated glass substrate \cite{GlassElastic}). We then zero-pad the signal at both beginning and end to obtain signals of uniform duration, and smooth with a $500$ MHz bandpass filter.

\section{Acknowledgements}

This work was supported by the the US National Science Foundation (grant no. CMMI-1333858), the US Army Research Office (grant no. W911NF-15-1-0030), and the University of Washington Royalty Research Foundation. The contribution by A.A.M. was supported by the National Science Foundation Grant No. CHE-1111557. M.H. acknowledges support from the National Science Foundation Graduate Research Fellowship Program under Grant No. DGE-1256082.

\section{Author Contributions}

M.H. performed the experiments. A.K. fabricated the samples. M.H. and M.A.G. conducted the data analysis. S.P.W. conducted the theoretical modeling. M.A.G. conceived the aluminum deposition as a method for interparticle stiffness tuning. All authors contributed to the development of concepts, the design of the experiments, and the writing of the paper.

%
%

\end{document}